\documentclass[letterpaper,USenglish]{lipics}
 
\usepackage{microtype}

\title{Improved reversible and quantum circuits for Karatsuba-based integer multiplication}
\titlerunning{Improved reversible and quantum circuits for Karatsuba-based integer multiplication}

\author{Alex Parent}
\affil{Institute for Quantum Computing\\University of Waterloo\\Waterloo, Canada\\
\texttt{alexparent@gmail.com}}
\author{Martin Roetteler}
\affil{Quantum Architectures and Computation Group\\Microsoft Research\\
Redmond, U.S.A.\\
\texttt{martinro@microsoft.com}}
\author{Michele Mosca}
\affil{Institute for Quantum Computing\\University of Waterloo\\Waterloo, Canada\\
\texttt{mmosca@iqc.ca}}

\authorrunning{A. Parent, M. Roetteler, and M. Mosca} 

\Copyright{Alex Parent, Martin Roetteler, Michele Mosca}

\subjclass{F.1.1 Models of Computation; F.2 Analysis of Algorithms and Problem Complexity}
\keywords{Quantum algorithms, reversible circuits, quantum circuits, integer multiplication, pebble games, Karatsuba's method.}

\usepackage{amsmath,amssymb,amsfonts,amsthm,amstext}

\usepackage{thmtools}

\usepackage{mathtools}
\mathtoolsset{centercolon} 

\usepackage{tikz}

\usepackage{cleveref}
\usepackage{tikz,circuitikz,tikzscale}

\usepackage{graphicx}
\usepackage{mathtools}
\usepackage{nicefrac}
\usepackage{cancel}
\usepackage{booktabs}
\usepackage{microtype}
\usepackage{pgfplots}

\usepackage[all]{hypcap}

\theoremstyle{definition}

\newcommand{\ket}[1]{|#1\rangle}

\newcommand{\F}{\mathbb{F}}

\newcommand{\Thm}[1]{\hyperref[thm:#1]{Theorem~\ref*{thm:#1}}}
\newcommand{\Lem}[1]{\hyperref[lem:#1]{Lemma~\ref*{lem:#1}}}
\newcommand{\Cor}[1]{\hyperref[cor:#1]{Corollary~\ref*{cor:#1}}}
\newcommand{\Def}[1]{\hyperref[def:#1]{Definition~\ref*{def:#1}}}
\newcommand{\Prop}[1]{\hyperref[prop:#1]{Prop.~\ref*{prop:#1}}}
\newcommand{\Prob}[1]{\hyperref[prob:#1]{Problem~\ref*{prob:#1}}}
\newcommand{\Fact}[1]{\hyperref[fact:#1]{Fact~\ref*{fact:#1}}}

\newcommand{\Sect}[1]{\hyperref[sect:#1]{Sect.~\ref*{sect:#1}}}
\newcommand{\Apx}[1]{\hyperref[apx:#1]{Appendix~\ref*{apx:#1}}}

\newcommand{\Fig}[1]{\hyperref[fig:#1]{Fig.~\ref*{fig:#1}}}
\newcommand{\Tab}[1]{\hyperref[tab:#1]{Table~\ref*{tab:#1}}}

\renewcommand{\copyrightline}{} 

\begin{document}

\maketitle

\begin{abstract}
Integer arithmetic is the underpinning of many quantum algorithms, with applications ranging from Shor's algorithm over HHL for matrix inversion to Hamiltonian simulation algorithms. A basic objective is to keep the required resources to implement arithmetic as low as possible. This applies in particular to the number of qubits required in the implementation as for the foreseeable future this number is expected to be small. We present a reversible circuit for integer multiplication that is inspired by Karatsuba's recursive method. The main improvement over circuits that have been previously reported in the literature is an asymptotic reduction of the amount of space required from $O(n^{1.585})$ to $O(n^{1.427})$. This improvement is obtained in exchange for a small constant increase in the number of operations by a factor less than $2$ and a small asymptotic increase in depth for the parallel version. The asymptotic improvement are obtained from analyzing pebble games on complete ternary trees. 
\end{abstract}

%
%

\section{Introduction} \label{sect:Intro}

Multiplication of integers is a fundamental operation on a classical computer. In quantum computing, integer multiplication is also an
important operation and indeed is at the core of what needs to be performed
in order to carry out Shor's algorithm for factoring integers \cite{Shor:94}. While much effort has been spent on optimizing the
arithmetic needed to implement Shor's algorithm---e.g., via constant
optimization \cite{SM:2012}, see also \cite{SM:2013}---the basic underlying
method for multiplication considered in most works is the simple school method
for multiplying integers that runs in time $O(n^2)$ elementary operations.
Elementary operations are here counted e.g. as the total number of Toffoli
gates, which form a universal gate set. Significantly less effort has been
spent on leveraging methods for fast multiplication which are well
known classically, e.g., Karatsuba's method and other recursive methods. 

Shor's factoring algorithm is special in that only multiplication by constants are required, which leads to significant simplifications in the circuits to implement Shor's algorithm \cite{Shor:94}. For more general period finding problems, e.g., Hallgren's algorithm \cite{Hallgren} and generalizations to computing the unit group in number fields of arbitrary degree \cite{EHKS} and to computing class numbers and the principal ideal problem \cite{Biasse}, more advanced arithmetic is required. This includes polynomial arithmetic which as a primitive building block requires integer multiplication $\ket{x,y,0} \mapsto \ket{x,y,xy}$ where inputs $x$ and $y$ can {\em both} be in superposition. 

Another example is the quantum algorithm for nonlinear structures \cite{CSV2007}: a full circuit level implementation of this algorithm will require the implementation of polynomial arithmetic over a finite field, which typically is reduced to integer arithmetic. Further examples where integer multiplication is a useful primitive is to implement a fast quantum Fourier transform: it was shown in \cite{CleveWatrous2000} that the computation of the Fourier transform can be reduced to integer multiplication, i.e., any fast algorithm for this problem gives rise to a quantum circuit for computing a Fourier transform on a quantum computer with the same time complexity. 

Finally, the implementation of arithmetic functions such as integer multiplication is an important primitive for quantum simulation algorithms \cite{BCK2015,BCCKS,Low}. Once a full gate level implementation of the quantum simulation algorithms is performed, arguably arithmetic operations are useful to implement the indexing functions of row- and column-computable matrices that appear in the decomposition of the Hamiltonian that is to be simulated. A similar reasoning applies to HHL type algorithms for matrix inversion \cite{HHL,Clader}, where the implementation of the underlying matrix may involve arithmetic operations such as integer multiplication for the computation of the entries. 

A simple approach to integer multiplication is to reduce it to addition in a straightforward way by using $n$ adders as in the familiar school method. 
If we let $Size(n)$ denotes the total size of a circuit---measured as the total number of Toffoli gates---where $n$ is the bit-size of the numbers to be multiplied. $Depth(n)$ denotes the depth of the circuit, allowing gates to be applied in parallel, and $Space(n)$ denotes the total space
requirements including input qubits, output qubits, and ancillas (i.e., qubits needed for intermediate scratch space), then the school method requires $Size(n)=Depth(n)=O(n^2)$ and $Space(n)=O(n)$. 

Classically, Karatsuba's algorithm allows to reduce the circuit size from $O(n^2)$ to $O(n^{\log_2 3})$ by recursively decomposing the problem for size $n$ into $3$ subproblems of size $n/2$. However, there is an issue with applying this algorithm to the quantum case: while it is still possible to obtain a size reduction to $Size(n)=O(n^{\log_2 3})$, in the straightforward way of circuitizing the recursion also the space complexity increases, so that overall $O(n^{\log_2 3})$ qubit are required. This was observed in the earlier work \cite{KRF:2006}, where also an improvement of the total depth to $O(n)$ was obtained, however, the number of qubits still scaled as $O(n^{\log_2 3})$. 

As quantum memory is a very scarce commodity and indeed early quantum computers are expected to only support a few hundred or perhaps thousands of logical qubits,
it is paramount to save space as much as possible. This leads to the question:

{\em Can recursions be leveraged on a quantum computer in such a way that the
space overhead does not grow as the total size of the circuit?  }

Or in a small variation of the above question: when considering the {\em
volume} of a quantum circuit computing the integer product of two $n$ bit
numbers, where volume is defined as the circuit depth $\times$ circuit width, is
it possible to compute this product in a volume that is strictly smaller than
$O(n^{1+\log_2 3})$ which was the previously best volume?

\newpage

{\bf Our results.} The results of \cite{KRF:2006} and the results derived in this paper can be compared as in the following table. Here ``parallel'' and ``sequential'' refer to different ways the recursion was unraveled in \cite{KRF:2006}, namely whether each of the $3$ circuits for subroutine calls to
problems of half size are arranged in parallel or are executed in sequence.

\bigskip

\begin{center}
\begin{tabular}{c@{\quad}c@{\quad}c}
\hline\hline
    Sequential \cite{KRF:2006} & Parallel \cite{KRF:2006} & This paper \\
    \midrule
    $\begin{aligned}
        Size(n) &= O(n^{\log_2 3})\\
        Depth(n) &= O(n^{\log_2 3}) \\
        Space(n) &= O(n^{\log_2 3}) 
    \end{aligned}$
	&
	$\begin{aligned}
        Size(n) &= O(n^{\log_2 3})\\
        Depth(n) &= O(n) \\
        Space(n) &= O(n^{\log_2 3}) 
    \end{aligned}$
    &
    $\begin{aligned}
        Size(n) &= O(n^{\log_2 3})\\
        Depth(n) &= O(n^{1.158}) \\
        Space(n) &= O(n^{1.427}) 
    \end{aligned}$\\
\hline\hline
\end{tabular}
\end{center}

\bigskip

Our main result is to give an affirmative answer to the question whether it is possible to implement recursions in less space than the circuit size dictates. More precisely, our implementation requires $O(n^{1.427})$ qubits which improves slightly over $O(n^{\log_2 3}) = O(n^{1.585})$, as recorded up to $3$ digits to the right of the decimal point in the last column of the table. For the total volume, defined as $Depth(n) \times Space(n)$, there is actually no advantage over \cite{KRF:2006} as it turns out that this quantity is asyptotically equal to $O(n^{1+\log_2 3})$. 

To achieve the bounds shown in the table, we apply a pebble game analysis of the recurrence structure of the Karatsuba algorithm. In this case the underlying graph that needs to be pebbled with as few pebbles as possible is a complete ternary tree. Perhaps surprisingly, even for seemingly simple graphs such as the complete $k$-ary trees, where $k=2$ or $k=3$, the optimal pebble game for a fixed number of pebbles seems not to be known. We provide a heuristic which allows to pebble the ternary tree corresponding to a bitsize of $n$ using $O(n \left(\frac{3}{2}\right)^{(\log_2 3)/(2 \log_2 3 - 1) \log_2(n)})=O(n^{1.427})$ pebbles. To the best of our knowledge, this is the first work that achieves an asymptotic improvement of the space complexity for integer multiplication while maintaining the $O(n^{\log_2 3})$ bound on the size of the quantum circuit. 

Besides the mentioned work \cite{KRF:2006} which
investigated Karatsuba-like circuits for integer multiplication, along similar
lines there is also work for the case of binary multiplication, i.e.,
multiplication over the finite field $\F_{2^n}$ \cite{KS:2015}. To analyze our algorithm we
use the framework of pebble games as introduced by Bennett \cite{Bennett:89} to
study space-time tradeoffs for reversible computations. The pebble games we
study are played on directed acylic graphs that have the structure of ternary
trees. In related work \cite{KSS:2016} pebbling of other classes of trees has
been considered, in particular that of complete binary trees.

\section{Preliminaries}

{\em The underlying gate model.} As with classical circuits, reversible functions can be constructed from
universal gate sets. It is known \cite{NC:2000} that the Toffoli gate which
maps $(x,y,z) \mapsto (x,y,z\oplus xy)$, together with the controlled-NOT gate
(CNOT) which maps $(x,y) \mapsto (x, x \oplus y)$ and the NOT gate which maps
$x \mapsto x \oplus 1$, is universal for reversible computation. When moving
from reversible to quantum computations, gate sets go beyond the set of
classical gates in that they allow to create so-called superposition of inputs.
For instance, popular choices of universal quantum gate sets are the so-called
Clifford+$T$ gate set and the Toffoli+Hadamard gate set. Universality in this
case means that it is possible to approximate any given target unitary
operation that we intend to execute on a quantum computer by a finite-length
sequence of operations over the given gate set. Herein the length of the sequence
typically scales as a polynomial in $\log(1/\varepsilon)$ where $\varepsilon$
is the target accuracy of the approximation, a result which has been established for the Clifford$+T$ gate set \cite{KMM:2013,Selinger:2015,RS:2016} as well as probabilistic variants thereof \cite{BRS:2015b,BRS:2015}. 

We point out that it is known that the Toffoli gate has an exact realization
over Clifford$+T$ \cite{NC:2000}, so all circuits for integer multiplication
presented in this paper can be exactly implemented over this gate set as well.
Furthermore, we refer the reader to \cite{Amy:2013} for more information about
the definition of $T$-depth and possible time-space tradeoffs for implementing
Toffoli gates and other reversible gates over the Clifford$+T$ gate set.

{\em Pebble games.} To study space-time tradeoffs in reversible circuit synthesis, Bennett \cite{Bennett:89} introduced reversible pebble games. This allow to explore ways to save on scratch space at the expense of recomputing intermediate results.

\begin{figure*}[htb]
\centering
\begin{tabular}{c@{\qquad}c@{\qquad}c}
\includegraphics[width=4cm]{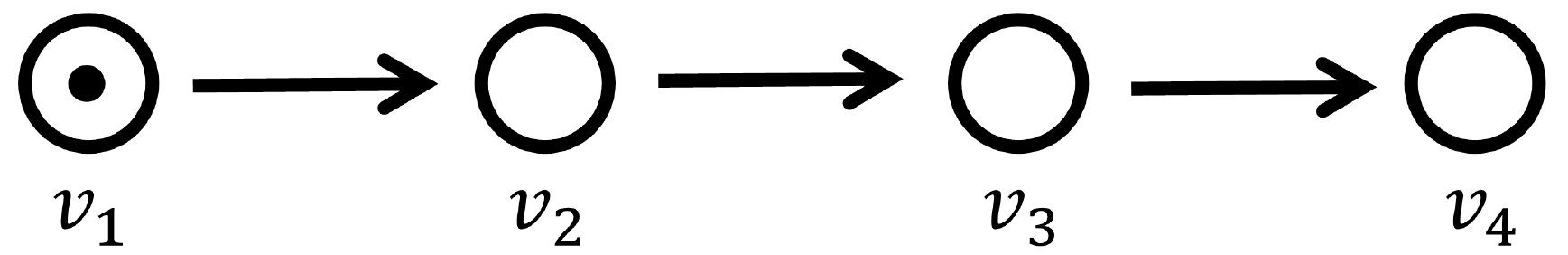} &
\includegraphics[width=4cm]{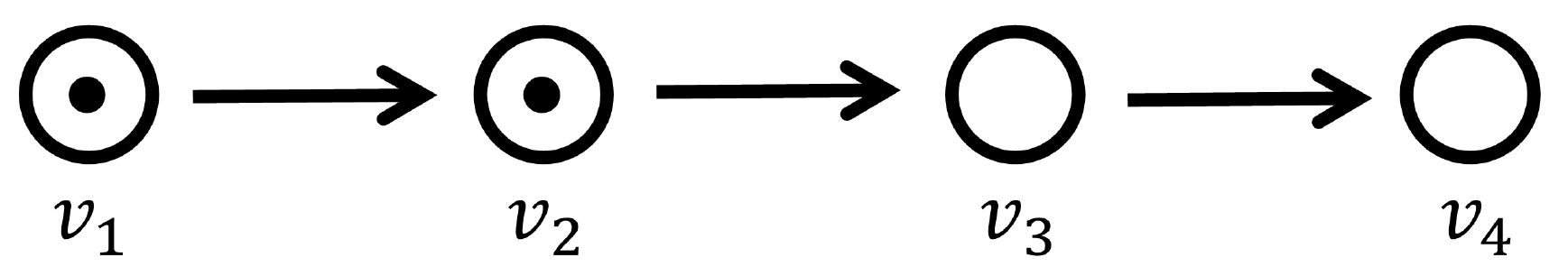} &
\includegraphics[width=4cm]{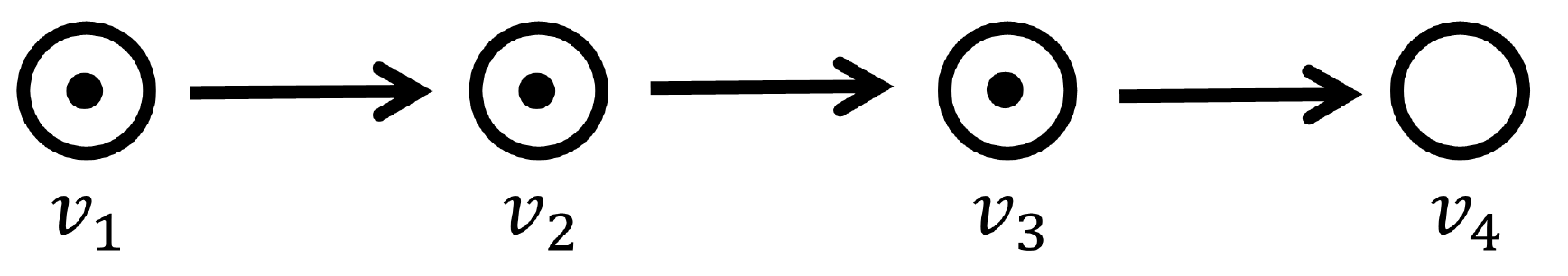}\\
(1) & (2) & (3) \\[3ex]
\includegraphics[width=4cm]{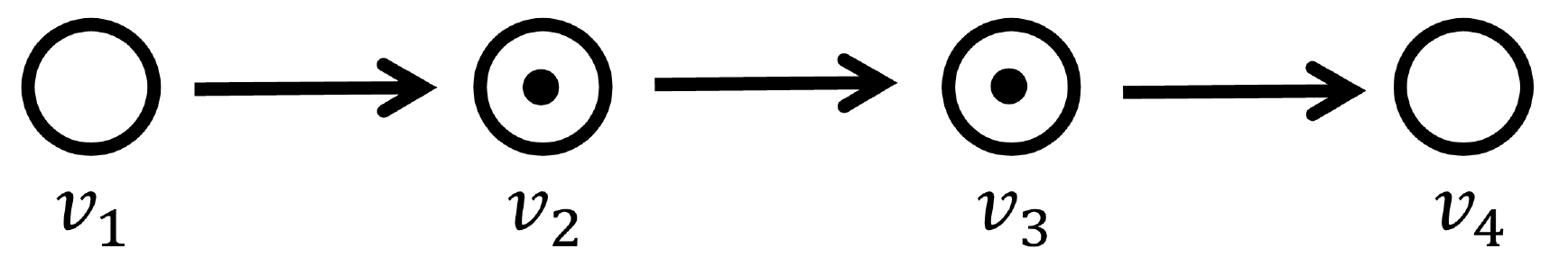} &
\includegraphics[width=4cm]{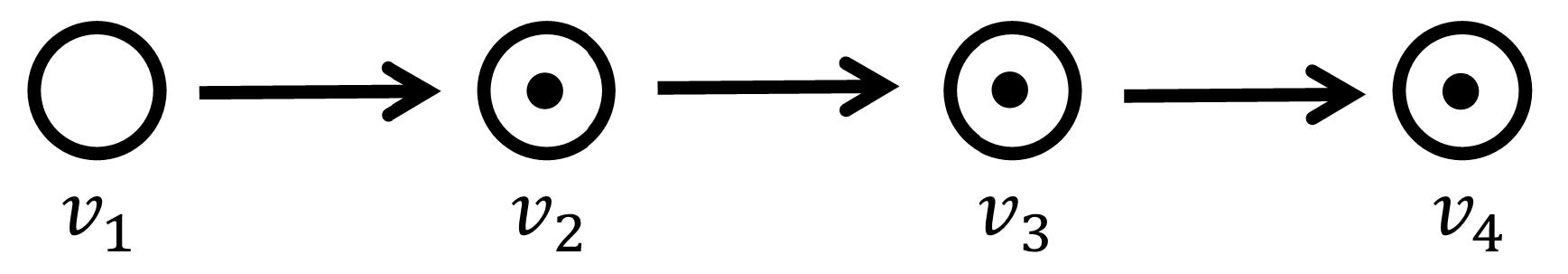} &
\includegraphics[width=4cm]{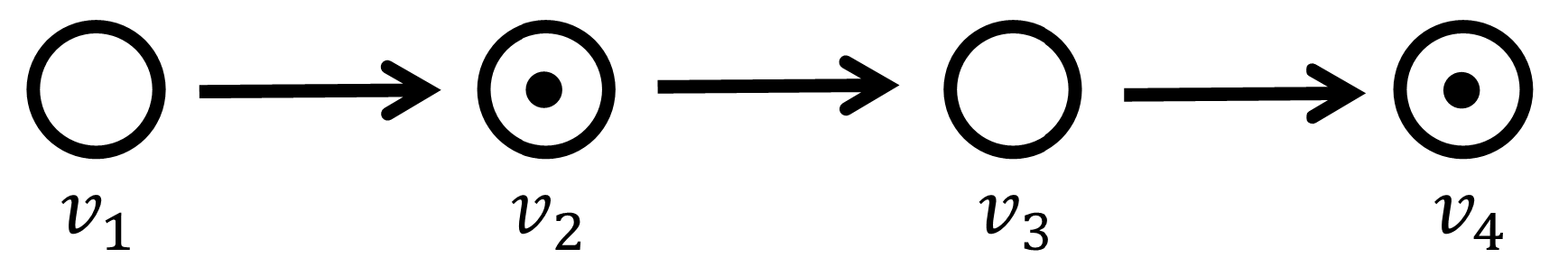}\\
(4) & (5) & (6) \\[3ex]
\includegraphics[width=4cm]{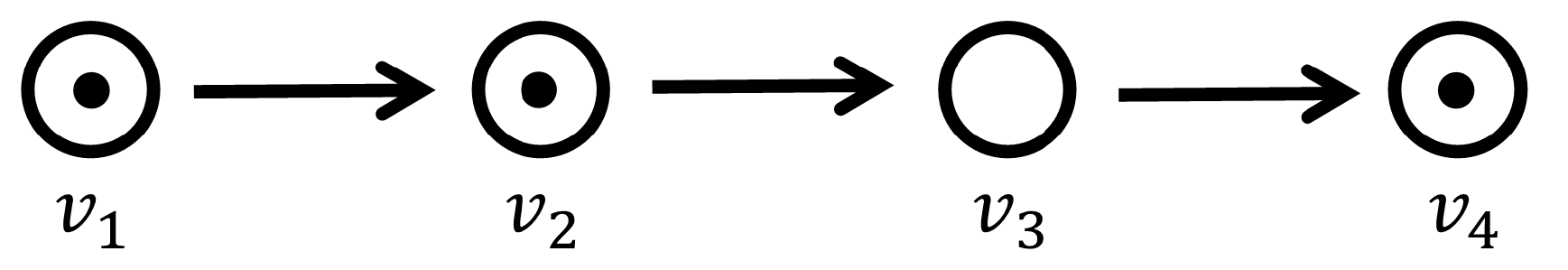} &
\includegraphics[width=4cm]{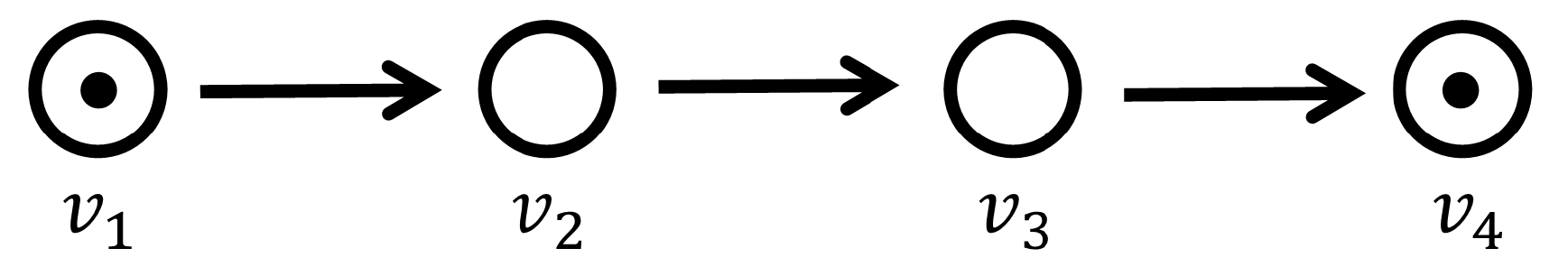} &
\includegraphics[width=4cm]{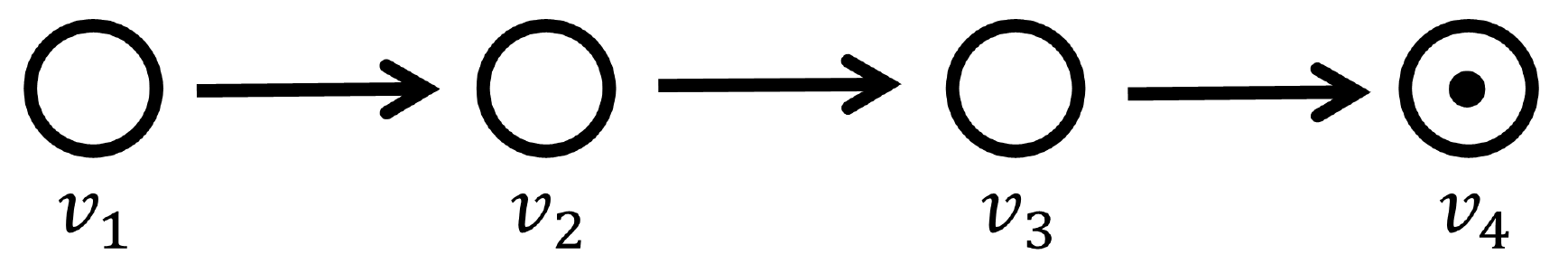}\\
(7) & (8) & (9) 
\end{tabular}
\medskip
\caption{\label{fig:pebblexample}
A pebble game played on a directed graph on $4$ vertices. If $4$ pebbles are available, one can simply proceed from left to right, pebbling one vertex at a time until the rightmost vertex is reached. After these $4$ steps, all pebbles except the one on the right are removed, requiring a total of $7$ steps. If only $3$ pebbles are available, the optimal strategy for this game requires $9$ moves which are shown in the subfigures (1) until (9).
}
\end{figure*}

A pebble game is defined on a directed acyclic graph $G=(V,E)$, where $V_{in} \subseteq V$ is a special subset of vertices of in-degree $0$, and $V_{out} \subseteq V$ is a subset of vertices of out-degree $0$. In each step of the game, a pebble can either be put or be removed on a vertex $v$, provided that for all $w\in V$ such that $(w,v)\in E$ already a pebble has been placed on $w$. Typically, a total bound $S \geq 0$ on the number of available pebbles is given. Vertices in $V_{in}$ can be pebbled at any time, provided enough pebbles remain. The task is to put a pebble on all vertices of $V_{out}$ and to do so in the minimal number of moves possible.  An example is given in Figure \ref{fig:pebblexample}. Here $V=\{v_1, v_2, v_3, v_4\}$, $V_{in}=\{v_1\}$, $V_{out}=\{v_4\}$. It turns out that the optimal strategy for $S=3$ requires $9$ steps and the corresponding moves are shown in subfigures (1) until (9). 

For a more formal treatment and further background information about pebble games we refer to \cite{Chan:2013}. If the graph on which the pebble game is played is a line, then the optimal pebbling strategies for a given space bound $S$ can be computed in practice quite well using dynamical programming \cite{Knill:95}. For general graphs, finding the optimal strategy is PSPACE complete \cite{Chan:2013}, i.e., it is unlikely to be solvable efficiently.

\begin{figure*}[htb]
\centering
\begin{tabular}{c@{\qquad}c@{\qquad}c}
\includegraphics[width=1cm]{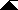} &
\includegraphics[width=3.5cm]{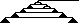} &
\includegraphics[width=7.5cm]{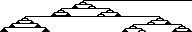} \\
(a) & (b) & (c)\\
\end{tabular}
\medskip
\caption{\label{fig:pebblegames}
Visualization of three different pebble strategies. (a) Bennett's strategy; (b) middle-ground heuristic strategy; (c) Lange-McKenzie-Tapp method.
}
\end{figure*}

In Figure \ref{fig:pebblegames} we display three different pebbling strategies that all succeed in computing a pebble game for the special case of linear graph, similar to one shown in Figure \ref{fig:pebblexample}, but for much larger number of vertices. In Figure \ref{fig:pebblegames} time is displayed from left to right, vertices are displayed vertically, with the vertex in $V_{in}$ on the bottom and the vertex in $V_{out}$ on top. The strategy shown in (a) corresponds to Bennett's compute-copy-uncompute method \cite{Bennett:73} where the time cost is linear. The strategy shown in (c) corresponds to the Lange-McKenzie-Tapp method \cite{LMT:2000} that resembles a fractal. In (b), a possible middle ground is shown, namely an incremental heuristic that first uses up as many pebbles as possible, then aggressively cleans up all bits except for the last bit, and the repeats the process until it ultimately runs out of pebbles. 

For a line graph with $|V|=n$, the Lange-McKenzie-Tapp strategy requires only $O(\log(n))$ pebbles and has an overall number of $O(n \log(n))$ steps, i.e., it is known that the line can be optimally pebbled in a number of steps that scales poynomially with the number of vertices. 

If the underlying graph $G$ is a complete binary tree on $n$ vertices such a polynomial bound is unfortunately not known. While it is known that the smallest number of pebbles required to pebble a binary tree of height $h$ is given by $S=\log(h)+\Theta(\log^*(h))$, where $\log^*$ denotes the iterated logarithm, to our knowledge the best upper bound on the number of steps is $n^{O(\log \log(n))}$, given in \cite{KSS:2016}. It is an open problem if a binary tree on $n$ vertices can be pebbled with a polynomial number of steps provided that only $S$ pebbles are available, where $S$ is as above. In this paper, we consider complete ternary trees as they arise naturally from the Karatsuba recursion. However, we do not strive for the optimal strategy and are content with a strategy that is good enough to give an asymptotic improvement. 

\section{Addition}

\begin{figure}[ht]\centering
\includegraphics[width=\linewidth]{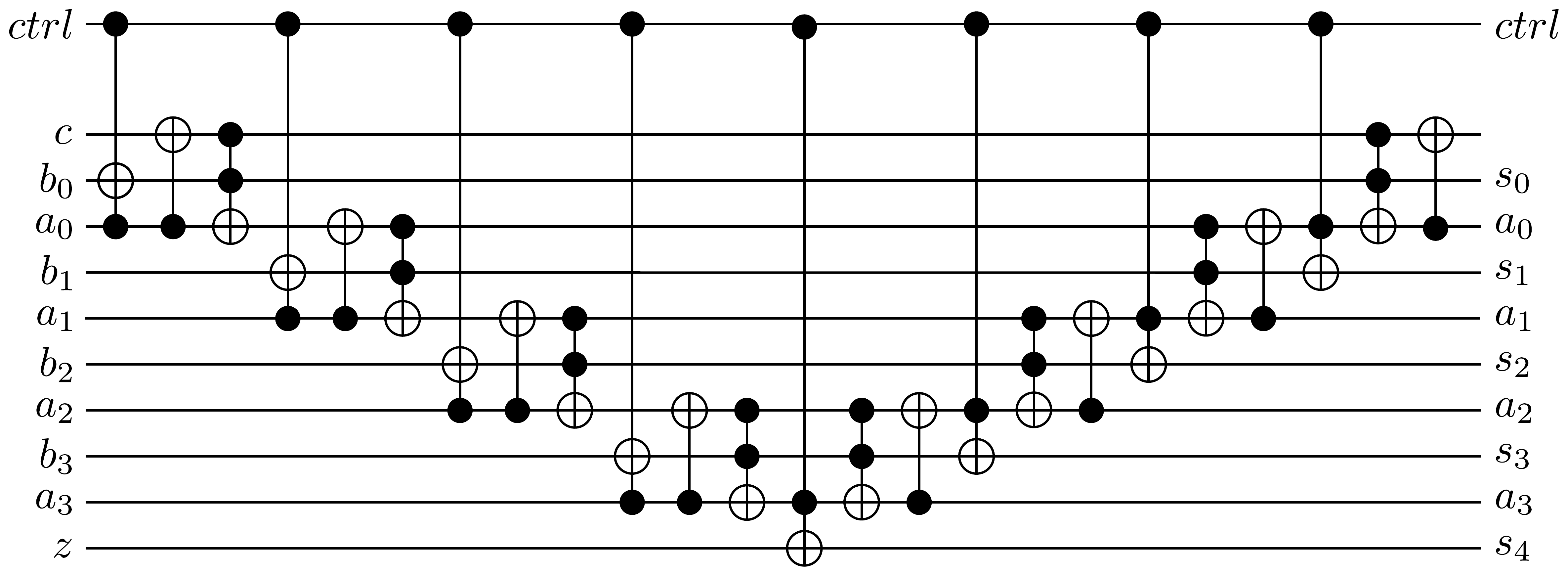}
\caption{Controlled ripple adder based on Cuccaro et al.~\cite{addCDK}.}
\label{fig:ctrlRipple}
\end{figure}

Circuits for multiplication of integers naturally rely on circuits to add
integers as subroutines, hence we first discuss circuits to perform addition.
The adder shown in Fig.~\ref{fig:ctrlRipple} is a circuit described in Cuccaro
et al.~\cite{addCDK} and forms the basis of simple multiplication circuits.

Note that not all the optimizations described in~\cite{addCDK} are
desirable in our context as we wish to minimize $T$ gates when adding controls to the overall circuit. It can be observed that that every Toffoli gate in the basic circuit given in~\cite{addCDK} shares its
controls with another.  We can therefore use ``directional'' Toffoli
gates~\cite{selinger2012}.  Each directional Toffoli uses four $T$-gates, requires
one ancilla and has a $T$-depth of one.  This circuit contains a total of $2n$ Toffoli gates and they are all in series.  The adder therefore has $8n$ $T$-Gates and a total $T$-depth of $2n$.

To implement a controlled adder we further note that not all gates in this circuit need be controlled: controlling a set of gates which if
removed would transform the circuit into the identity is sufficient. In the
case of the in-place adder the MAJ and UMA subcircuits that were introduced in
\cite{addCDK} can be made to cancel by removing one gate each.
\Cref{fig:ctrlRipple} shows the resulting circuit. The circuit has a total number of $4n$ Toffoli gates, all of which are in series. Therefore, the total $T$-count of the controlled adder is $16n$ and the total $T$-depth is $4n$. 

\begin{figure}[ht]\centering 
\includegraphics[width=0.6\linewidth]{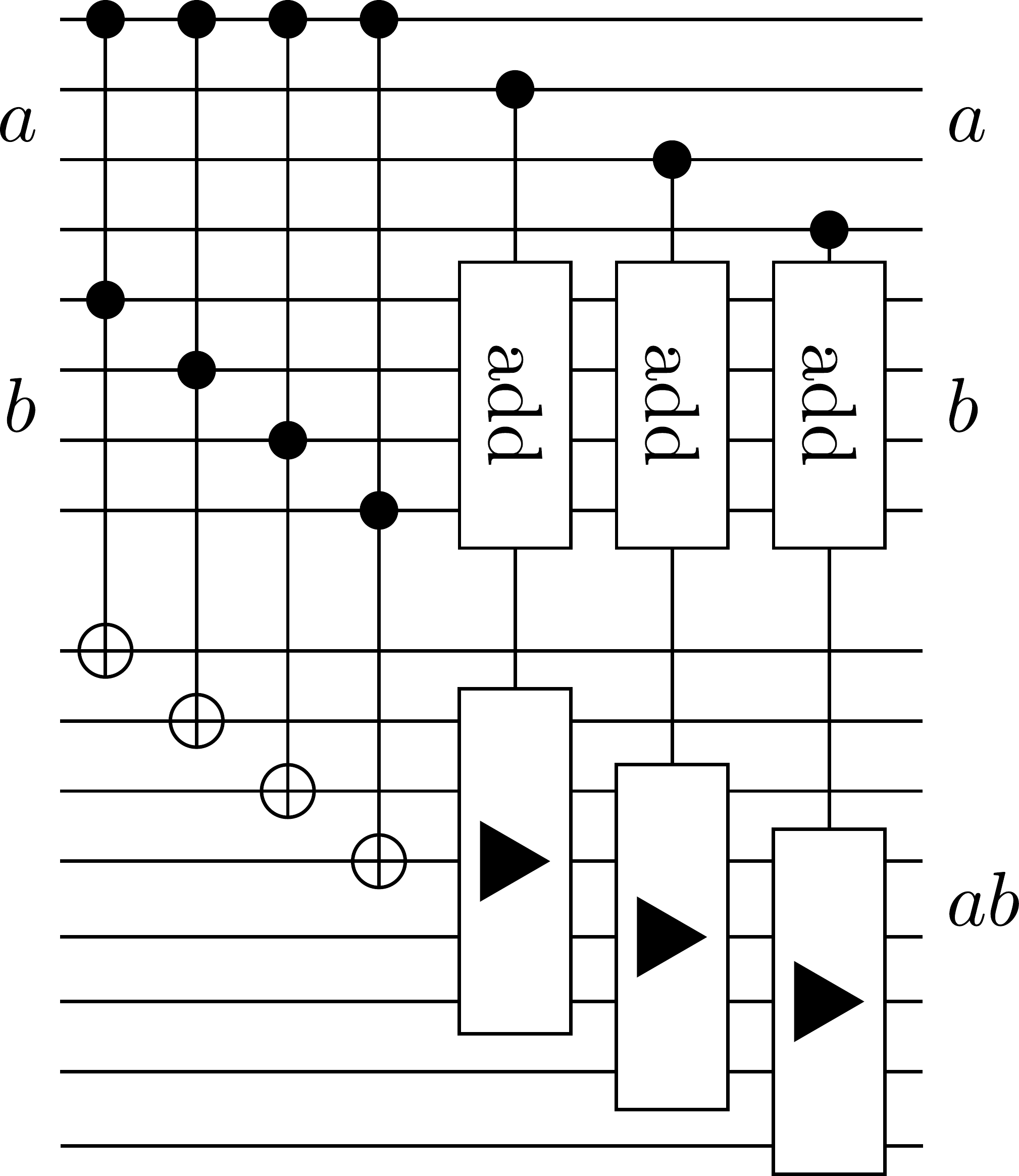}

\caption{Controlled addition multiplier. In the above circuit notation the
triangle designates the modified bits in the adder. The circuit consists of a sequence of
controlled additions as in Fig.~\ref{fig:ctrlRipple} with the exception of the
first block which can be replaced by a cascade of Toffoli gates as the ancilla
qubits at the bottom are initialized in the zero state. The total gate count
scales asymptotically as $O(n^2)$.}

\label{fig:multAdd}
\end{figure}

A simple $O(n^2)$ implementation of multiplication as a controlled addition circuit
is shown in Fig.~\ref{fig:multAdd}.  Given two numbers as bit strings $a$ and $b$
their product can be found by repeatedly shifting forward by one and adding $b$
to the result controlled on the next bit in $a$. The overall circuit is an out-of-place multiplier that uses only $1$
additional ancilla for the adder circuits.

This circuit takes $n$ Toffoli gates to copy down the initial value.  It then
uses $n-1$ controlled in place addition circuits to produce the final value.
If we define $A^{ctrl}_n$ to be the Toffoli count for a controlled adder of
size $n$ we get $M_n = n + (n-1)A^{ctrl}_n$, where $M_n$ is the gate count for
a controlled addition based multiplication circuit of size $n$.  We know from the above discussion that the
controlled addition circuit uses $4n$ Toffoli gates. This yields a total Toffoli count of the integer multiplication of
\begin{equation} \label{eq:caddtoff}
    M_n = 4n^2 -3n,
\end{equation}
and a space complexity that scales linear with the number of qubits. 

The rest of the paper will consider methods to reduce this total gate count to
$O(n^{\log_2 3})$ while improving the amount of ancillas that are required to do
so when compared to prior approaches.

\section{Reversible Karatsuba multiplier\label{sec:kara}}

The following reversible algorithm for Karatsuba improves upon previous
work~\cite{KRF:2006}.  It does this primarily by using in place addition to
minimize garbage growth at each level.  It also attempts to choose optimal
splits instead of dividing the number in half at each step, This is helpful
when the integer size is not a power of 2. Further an asymptotic
improvement in space use (yielding as well an asymptotic improvement in the
space-time product), is shown by using pebble games in the analysis.

Let $n\geq 1$ and let $x$ and $y$ be $n$-bit integers.  The well-known
Karatsuba~\cite{KO:62} algorithm is based on the observation that by
writing $x=x_1 2^{\lceil n/2\rceil}+x_0$ and $y=y_1 2^{\lceil n/2\rceil
}+y_0$ the product $xy$ can be evaluated as $xy=2^n A + 2^{\lceil n/2
\rceil} B + C$, where
\begin{align*}
  A &= x_1 y_1, \\
  B &= (x_0+x_1)(y_0+y_1) - x_0 y_0 - x_1 y_1,\\
  C &= x_0 y_0.
\end{align*}

Note that computation of $A$, $B$, and $C$ only requires multiplication of
integers that have bits size $n/2$, i.e., half the bit size of $x$ and $y$. The
final addition is carried out as the addition of $n$ bit integers.

\begin{figure}[hbt]
  \capstart
  \centering
  \includegraphics[width=0.9\linewidth]{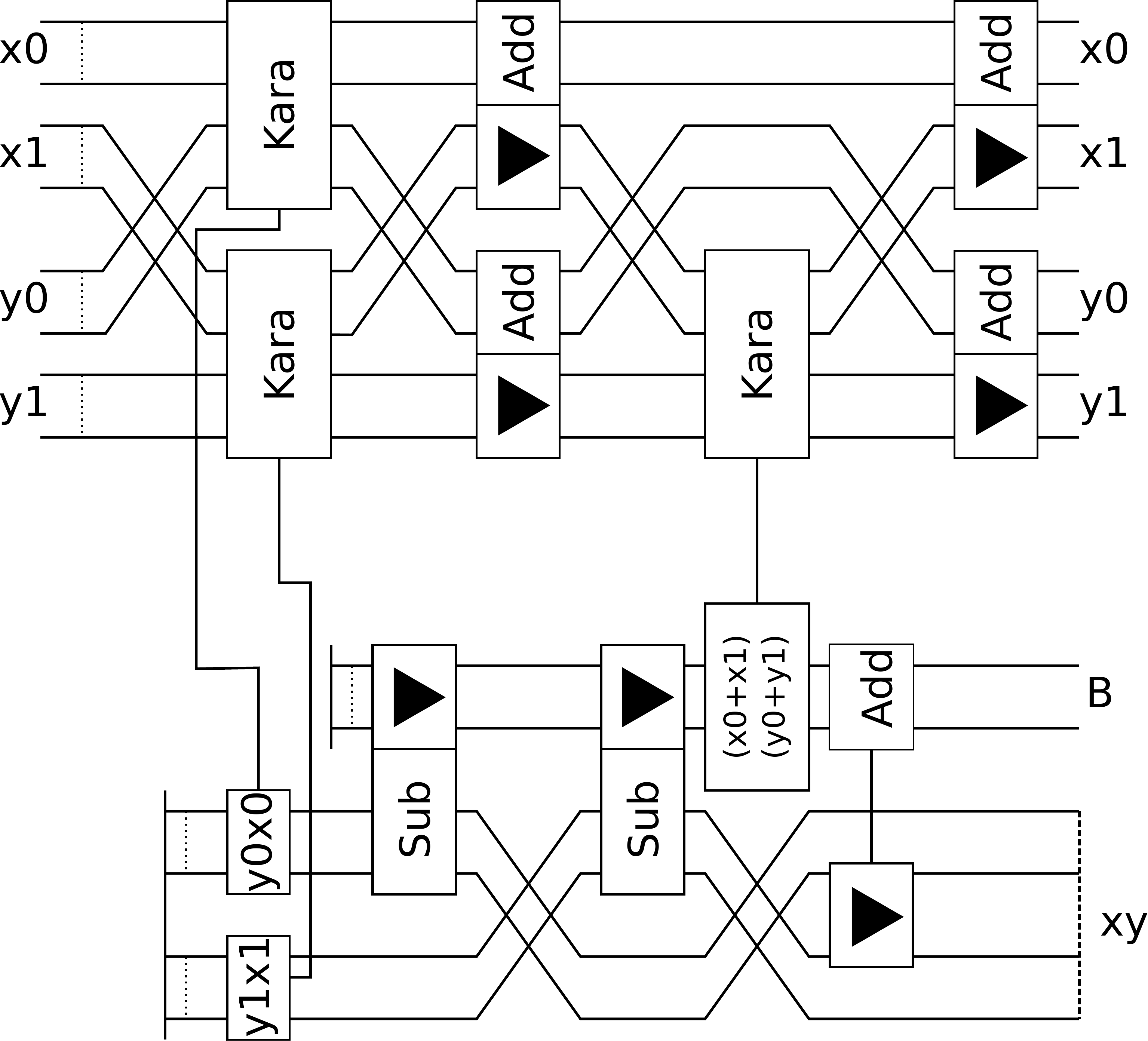}

  \caption{Karatsuba multiplication circuit. Besides the output (denoted
``$xy$'') this circuit outputs also the intermediate result ``$B$'' as in the
Karatsuba recursion $xy=2^n A+2^{\lfloor n/2 \rfloor} B + C$ mentioned in the
text. In order to remove $B$, we copy out the result ``$xy$'' and run the
circuit backward. The main contribution of this paper is an analysis on when to
perform this uncomputation as a function of the level of the recursion. Note
that the final two adders return the inputs to their original state in order to
save space. These adders can be removed at the cost of additional garbage
bits.}

  \label{fig:kara2}
 \end{figure}
\subsection{Analysis}

\begin{figure}[hbt]
\capstart
\includegraphics[width=0.9\linewidth]{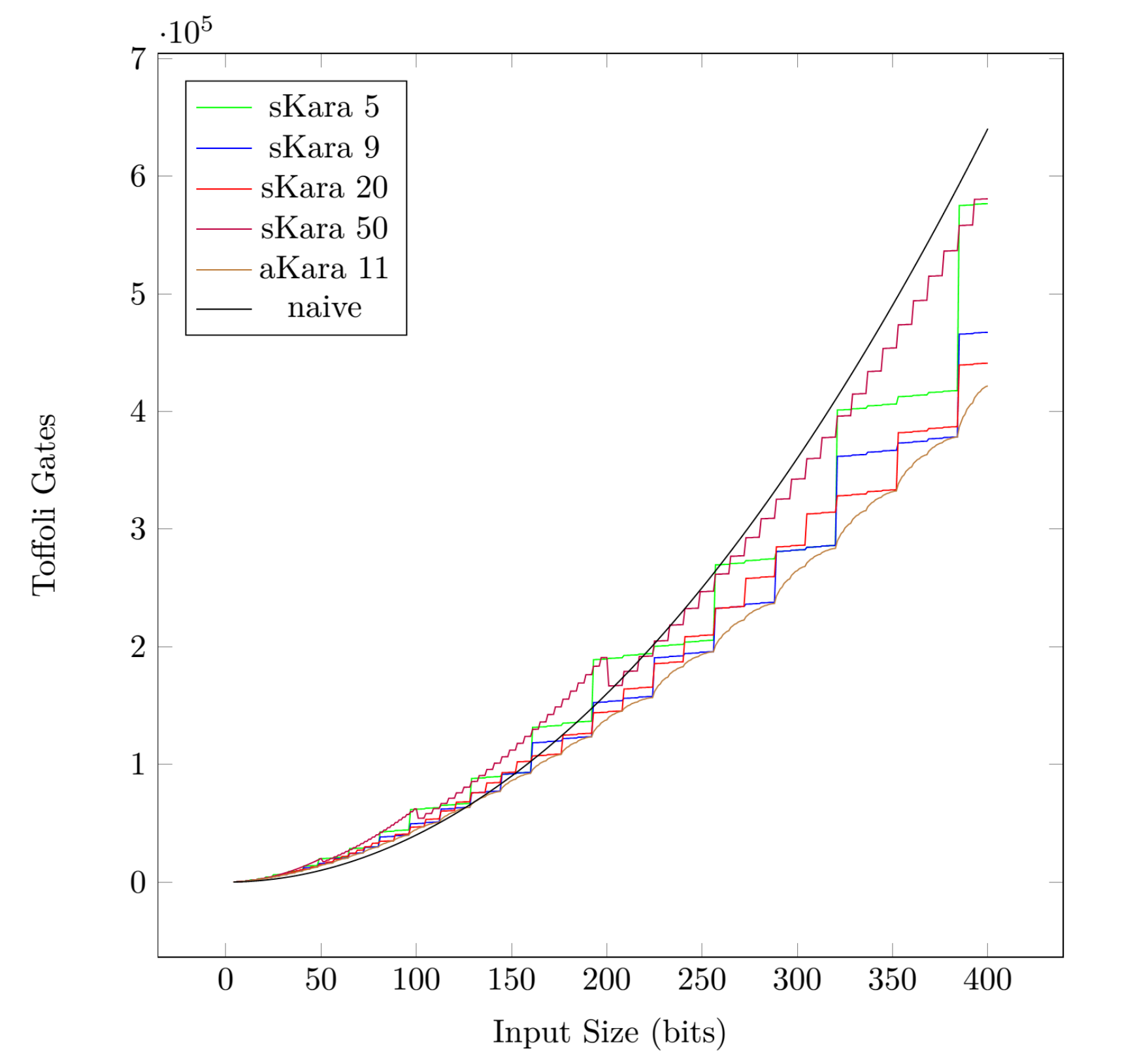}

\caption{Plot of circuit sizes versus input size for various various Karatsuba
cutoffs.  The Legend shows the implementation (skara for the simple version and
aKara for the adaptive cutoff) as well as a number indicating the cutoff size. For instance for a bit-size of $n=400$ the naive method requires about $400 n^2 = 640,000$ Toffoli gates, whereas the best strategy \texttt{aKara11} found by our search requires only about $422,000$ Toffoli gates.  
}

\label{fig:cutoff}
\end{figure}

  Note that the cost for the computation of $A$, $B$, and $C$ are $3$
  multiplications and four additions.  Note further that the additions to
  compose the final result do not have to be carried out as the bit
  representation of $xy$ is the concatenation of the bit representations of
  $A$, $B$, and $C$.  For $m\geq 1$, let $M^{g}_m$ denote the Toffoli cost
  of a circuit that multiplies $m$-bit inputs $x$ and $y$ using ancillas,
  i.e., a circuit that maps $(x, y, 0, 0) \mapsto (x, y, g(x,y), xy)$,
  where $xy$ is a $2m$-bit output, and $g(x,y)$ is an garbage output on
  $k\geq 1$ bits.  Furthermore, denote by $A_m$ the cost for an (in-place)
  adder of two $m$-bit numbers.  It is known that $A_m$ can be bounded by
  at most $2m$ Toffoli gates. Let $K_n$ denote the number of Toffoli gates
  that arise in the quantum Karatsuba algorithm (See Fig.~\ref{fig:kara2}).
  The outputs of one step of the recursion are $x_0$,$x_1$, $y_0$, $y_1$,
  $x_0y_0$, $x_1 y_1$, and $xy$.  It is easy to see that allowing garbage,
  $K_n^g$ can be implemented using $3$ multipliers of half the bit size,
  $4$ in-place adders of size $n$ and $4$ in place adders of size $n/2$
  (note the subtracters are just reversed adders).  The base case is a
  multiplier for two one-bit numbers which can be done with one Toffoli
  gate, i.e., $K_1^{g}=1$.  We obtain the following recursion:
  \begin{align}
    K_n^{g} = 3 K_{n/2}^{g} + 4 \left(A_{n} + A_{n/2}\right); \quad K_1^{g}=1.
  \end{align}
  For the overall clean implementation of the Karatsuba algorithm we first
  run this circuit forward, copy out the final result using $n$ CNOTs, and
  then run the whole circuit backward.  This leads to an overall cost of
  $K_n = 2 K_n^{g}$ and $n$ CNOTs.  For the moment we focus on the Toffoli
  cost only.  By expansion we obtain that:
  \begin{align}
    K_n^{g} &= 3^{\log_2(n)} K_1^{g} + 4\left( A_{n}+ A_{n/2} \right) + 12\left( A_{n/2}+ A_{n/4} \right)\notag\\
            & + \ldots + 4 \cdot 3^{\log_2(n)-1} \left(A_2 + A_1\right).
  \end{align}
  Using that the Toffoli cost of $A_{n/2^i}$ is $2(n/2^i)$, we obtain for
  the overall Toffoli cost the following bound:
  \begin{align}
    K_n &= 2\left(3^{\log_2 n } + 4 \sum_{i=0}^{\log_2 n - 1} 3^i 2(3n/2^i)\right)\notag\\
        &= 2n^{\log_2 3} + 48\,n \left(\frac{1- (3/2)^{\log_2 n}}{1-3/2}\right) \notag\\
        &= 2n^{\log_2 3} + 96\,n \left((3/2)^{\log_2{n}} -1\right) \leq 98\, n^{\log_2{3}}.
  \end{align}
This bound can be improved by replacing the recursive call to Karatsuba
  with naive multiplication once a certain cutoff has been reached.
  In Fig.~\ref{fig:cutoff} we provide a comparison of various cutoff values
  (the naive method based on eq.~(\ref{eq:caddtoff}) is also plotted for reference).

\begin{figure}[hbt]
\capstart
\includegraphics[width=0.9\linewidth]{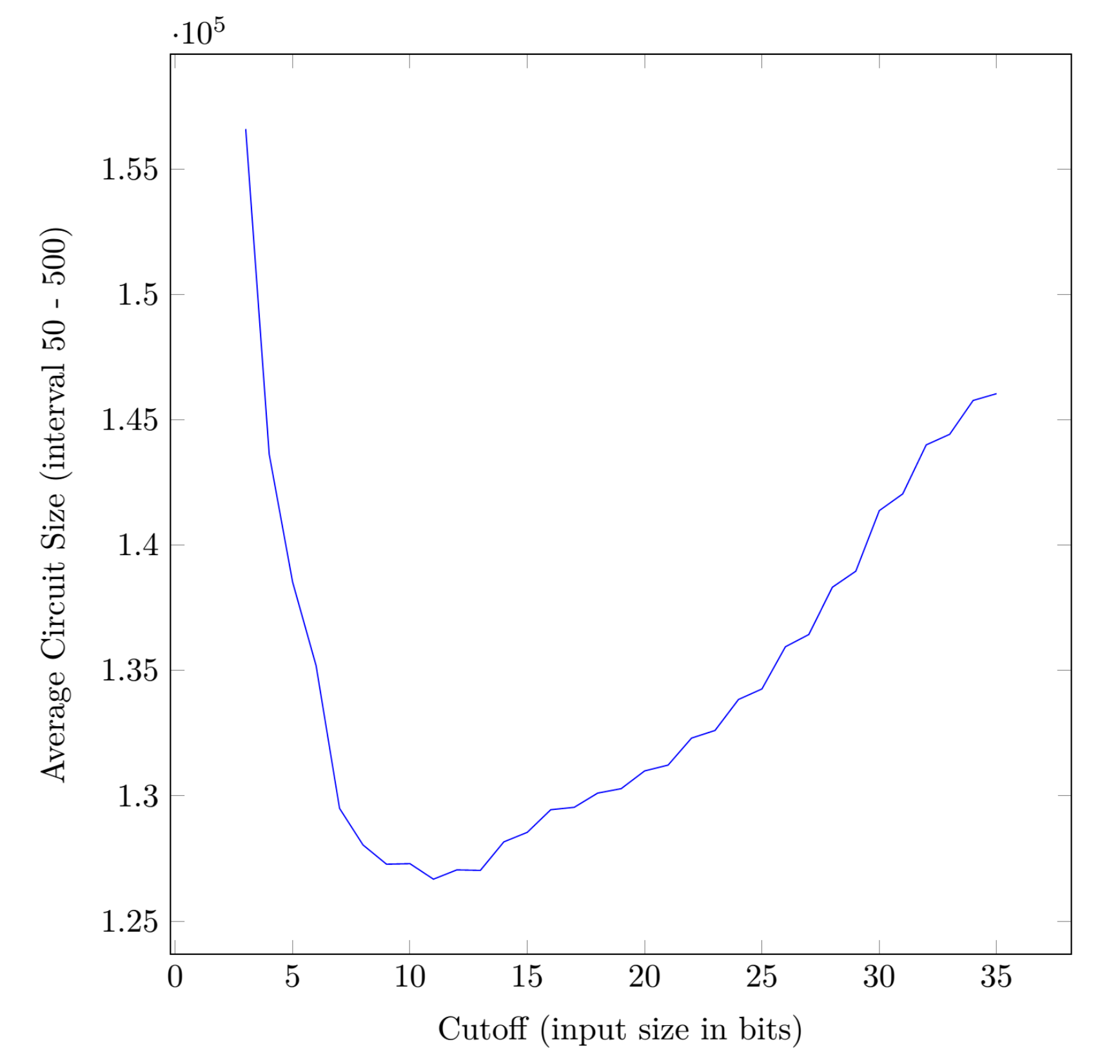}
\caption{Average circuit size over the interval 50-500 for various cutoff values.}
\label{fig:cutoffs}
\end{figure}

Another way to improve this algorithm is to attempt to choose more
intelligent splits rather than always splitting the inputs in half at
each level.  This is important because the bit length of the numbers we
are adding together may not be a power of two so dividing the input in
two at each level might not be optimal.  In Fig.~\ref{fig:cutoff} the line
plotted as \texttt{aKara11} shows the result of using the optimal splits at
each level.  These were found by a simple dynamic program which evaluated
the total gate size for every possible split at every level and chose the
optimal ones. Using these methods we find an optimal cutoff value of 11 (see Fig.~\ref{fig:cutoffs}).

\begin{figure}[hbt]
\capstart
\includegraphics[width=0.9\linewidth]{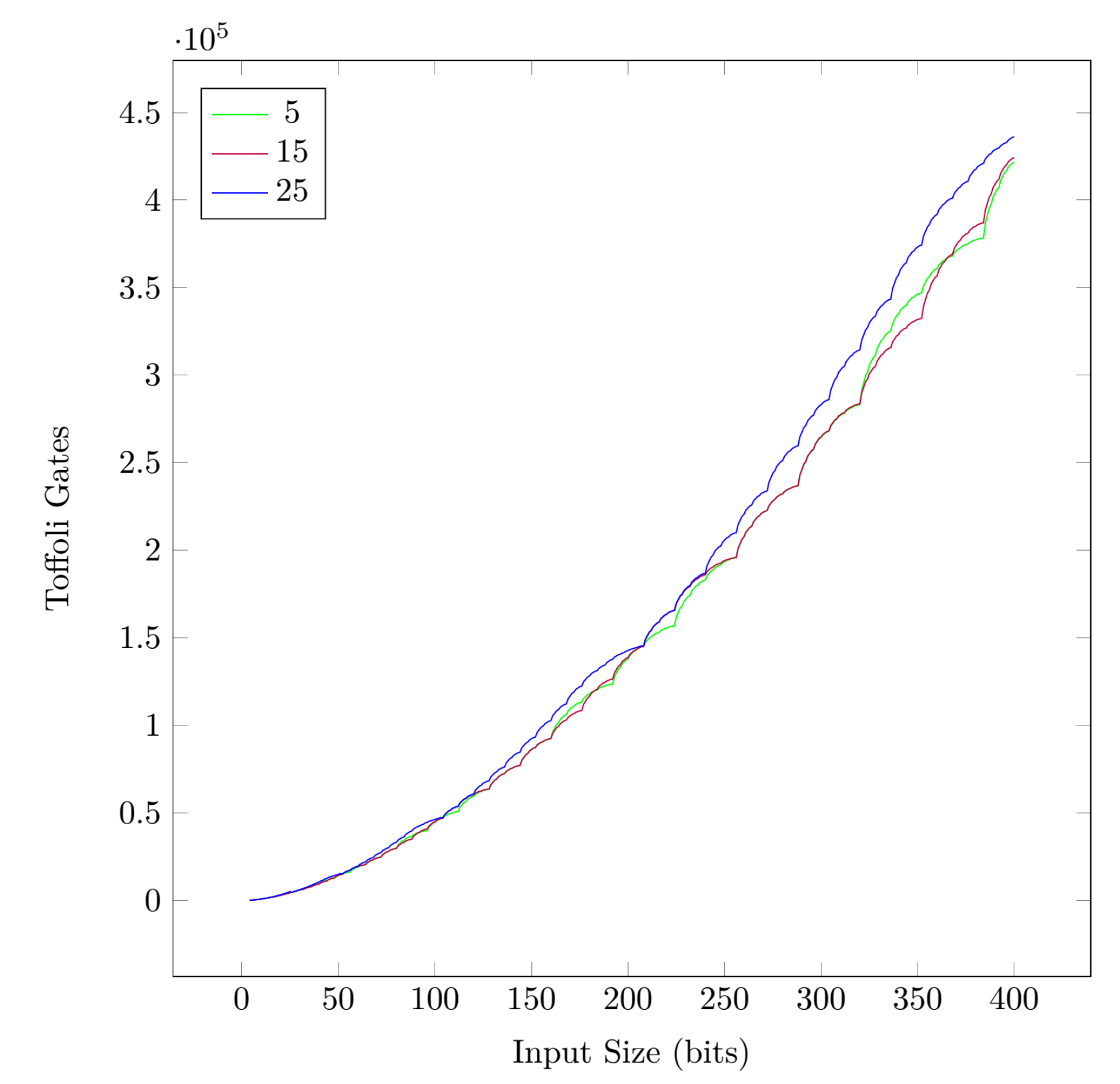}
\caption{Comparison of various choices for adaptive cutoffs.}
\label{fig:aKara}
\end{figure}

\subsection{Time-space tradeoffs}

\begin{figure}[hbt]
\capstart
\includegraphics[width=0.9\linewidth]{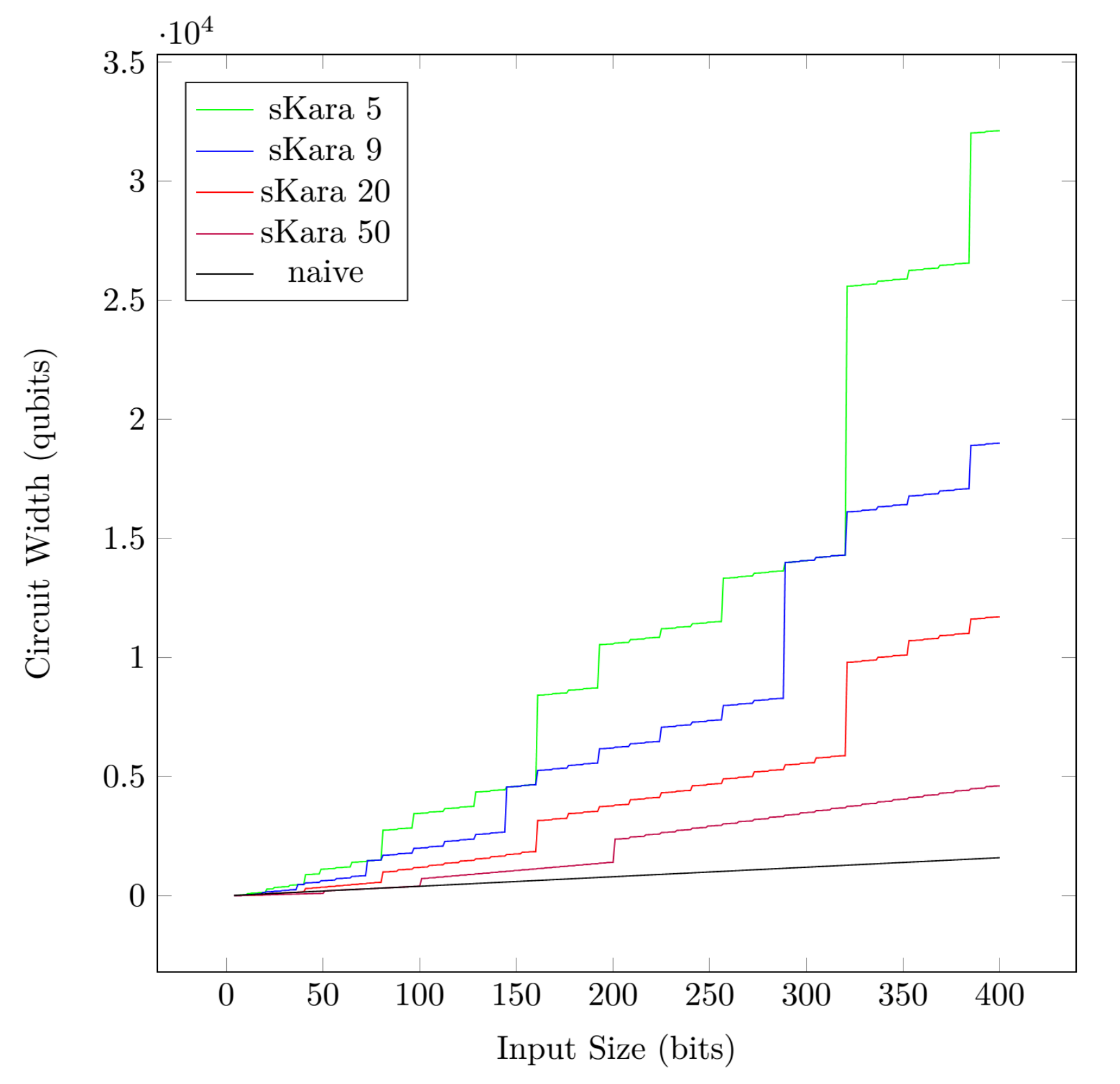}
\caption{Qubits used versus input size for various Karatsuba cutoffs.}
\label{fig:size}

\end{figure}

 We see in Figs.~\ref{fig:cutoff} and \ref{fig:size} that there are trade-offs available
 between circuits size and gate count available by changing the cutoff
 value.  A higher cutoff value results in a larger naive
 multiplication circuits which are much more space efficient.

\begin{figure}[hbt]
\capstart
\includegraphics[width=0.9\linewidth]{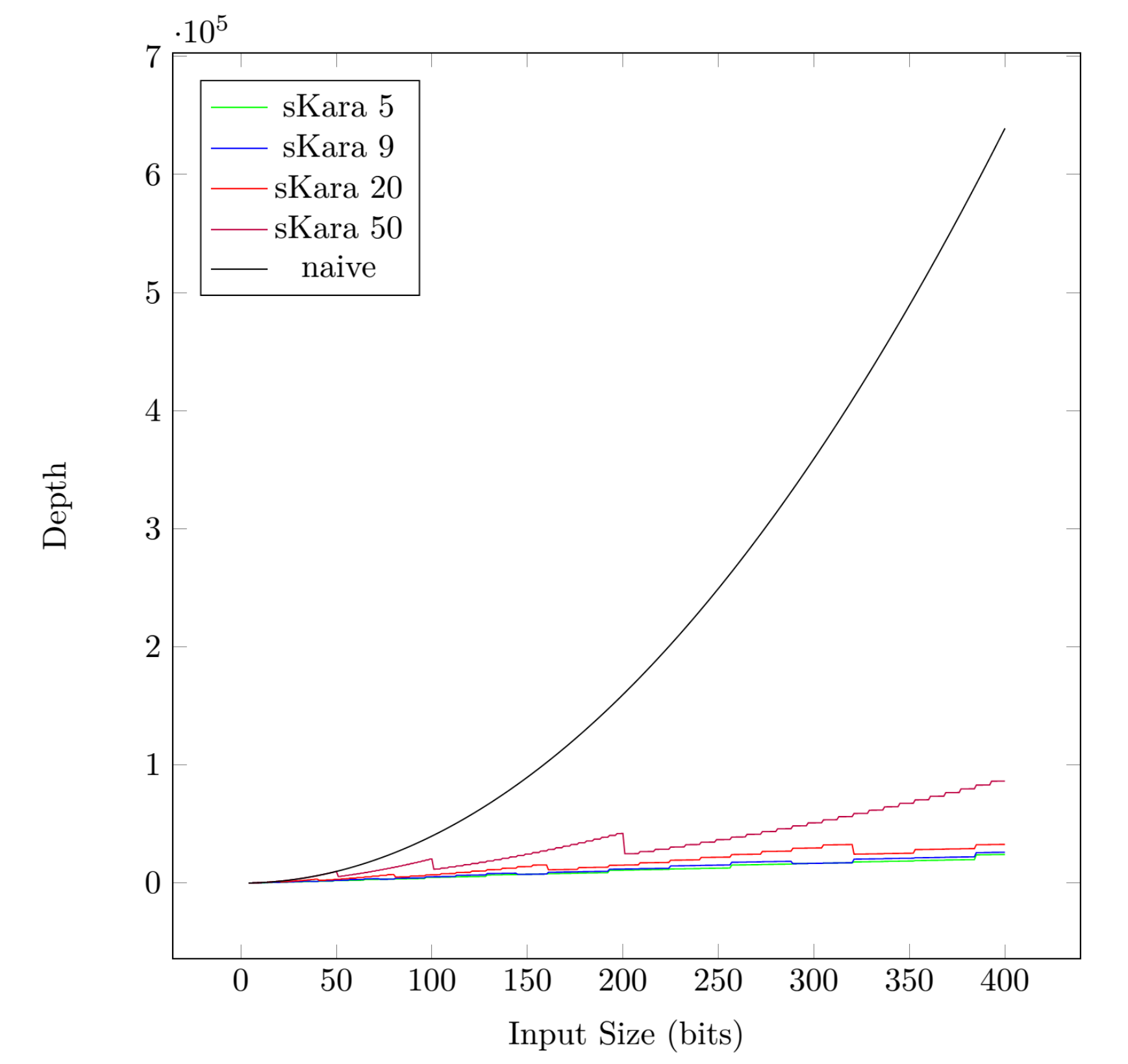}
\caption{Toffoli depth versus input size for various Karatsuba cutoffs.}
\label{fig:depth}
\end{figure}

\begin{figure}[hbt]
    \capstart
    \centering
    \includegraphics[width=0.8\linewidth]{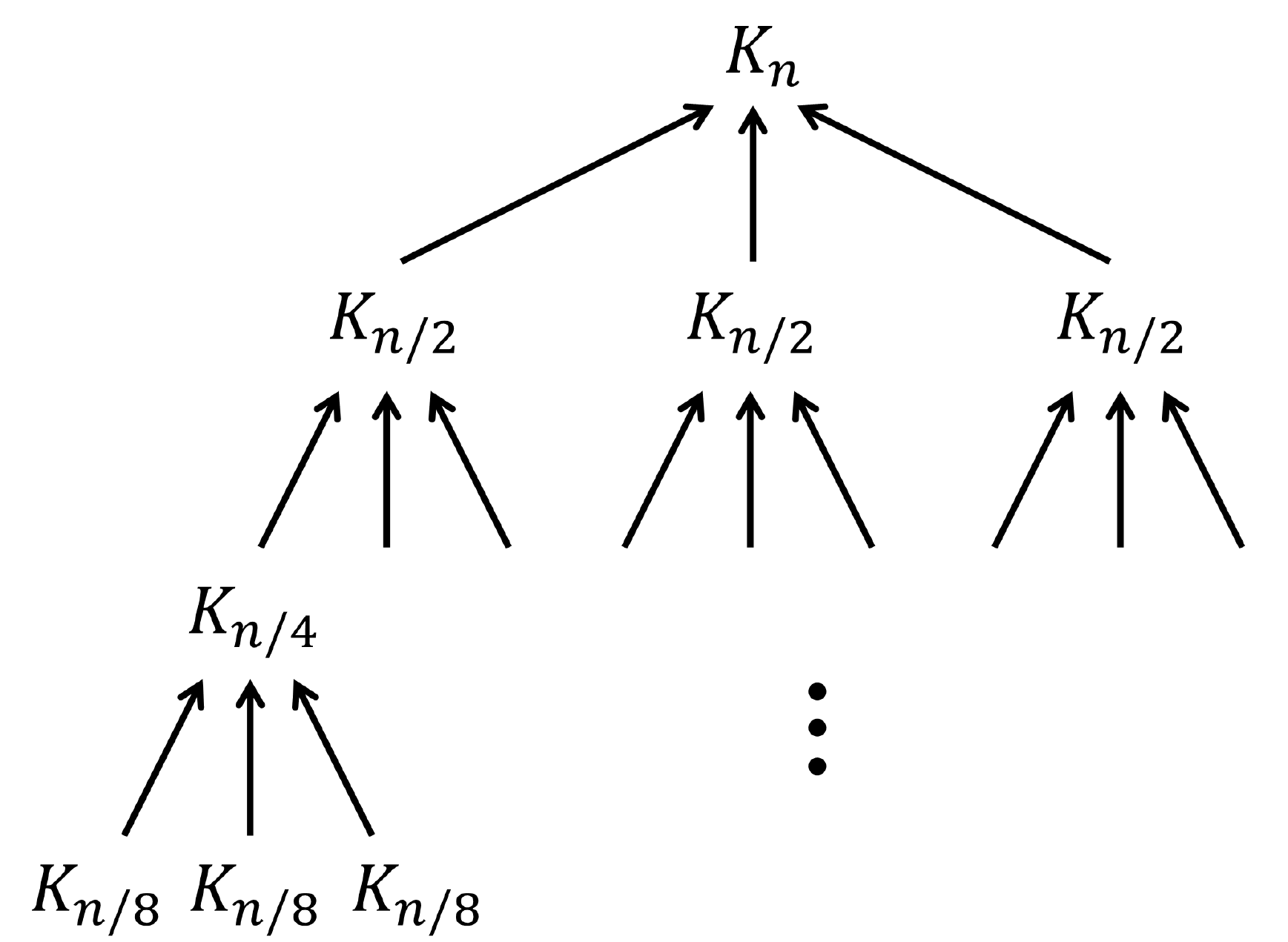}

    \caption{Structure of a pebble game for recursively implementing the Karatsuba circuit. Here $K_i$ for $i=1, 2, \ldots, n$ stands for the problem at level $i$, i.e., a problem with input-size $i$ bits.}

    \label{fig:kara-mdd}
\end{figure}

The reversible pebble game may be used to gain an asymptotic improvement in the
space required to implement this algorithm. Note the tree structure of the
recursive dependencies shown in Fig.~\ref{fig:kara-mdd}.  We find a level such
that the size of each node's subtree is approximately equal to the size of the
sum of all nodes at that level and above. Then for each node at that level in
sequence compute the node and uncompute all nodes below it.

For the Karatsuba circuit on input of size $n$ at a level $x$ in the tree there
are $3^x$ nodes of size $2^{-x}n$ for a total cost of
\[
n\left(\frac{3}{2}\right)^x.
\]

So the total cost of the full tree is given by
\[
    n\sum_{i=0}^{N} \left(\frac{3}{2}\right)^i,
\]
where $N=\log_2 n$. To pebble the underlying ternary tree, we would like to break the tree into approximately equal sized subtrees at
some level.  Each tree at that level will be computed then uncomputed
leaving only the top node. To minimize space we will choose the size of
these subtrees to be approximately equal to the remaining size of the tree
above them. In order to find the height $k$ of such a tree we set:

\[
    \sum_{i=0}^{N-k-1} \left(\frac{3}{2}\right)^i = \frac{1}{2^{N-k}}\sum_{i=0}^{k-1} \left(\frac{3}{2}\right)^i.
\]

Since this is a geometric series we can use the identity $\sum_{k=0}^{n-1} r^k
= \frac{1-r^n}{1-r}$ which holds for all $r$ and obtain

\begin{align*}
    \frac{1- \nicefrac{3}{2}^{N-k}}{1 - \nicefrac{3}{2}} &= \frac{1}{2^{N-k}}\frac{1- \nicefrac{3}{2}^{k}}{1 - \nicefrac{3}{2}}.
\end{align*}
Rearranging terms, we obtain
\begin{align*}
    1- \nicefrac{3}{2}^{N-k} &= 2^{k-N} - \frac{3^k}{2^N}.
\end{align*}
Since $k\leq N$ and since we want that $\nicefrac{3}{2}^{N-k} \geq
\frac{3^k}{2^N}$ a simple calculation shows that this will be the case for $k
\leq \frac{N}{ 2- \frac{\log 2}{\log 3}} = 0.731N$. The total space use without
this optimization can be calculated as
\[
    n\sum_{k=0}^{\log_2 n - 1} \left(\frac{3}{2}\right)^k = n\frac{1-(\nicefrac{3}{2})^{\log_2 n}}{1-\nicefrac{3}{2}}.
\]
This gives space use of $O(n(\nicefrac{3}{2})^{\log_2 n}))$ which is
equivalent to $O(n^{\log_2 3})$ or approximately $O(n^{1.585})$. Using the above optimization we get space usage that can be bounded by
\[
    O\left(n\left(\frac{3}{2}\right)^{\left({\frac{\log 3}{ 2\log 3 - \log 2}\log_2 n}\right)}\right) \approx O(n^{1.427}).
\]
To find the depth of the circuit note that each node at level $k$ must be
computed sequentially. At level $k$ the number of trees is
\[
    3^{\left(1-\frac{\log 3}{ 2\log 3 - \log 2}\right)\log_2 n}.
\]
Each tree is of depth
\[
    \frac{n}{2^{1-\frac{\log 3}{ 2\log 3 - \log 2}}}.
\]
This gives an overall depth for computing the $k$ level of
\[
    n\left(\frac{3}{2}\right)^{\left(1-\frac{\log 3}{ 2\log 3 - \log 2}\right)\log_2 n} \approx n^{1.158}.
\]
Overall, we get a space-depth volume of our circuit that scales as $n^{1+\log_2 3}$.

\subsection{Generalization to other recursions}

Assume that we are given a function with input size $n$ which splits a problem
into a total of $a$ subproblems of size $\nicefrac{n}{b}$ where the total cost to subdivide and recombine is $O(n)$. Then the overall work to compute the function for a problem of size $n$ is given by:

\[
    n\sum_{i=0}^{N} \left(\frac{a}{b}\right)^i.
\]

Solving as above we have:
\[
    k \leq \frac{\log_b n}{ 2- \frac{\log b}{\log a}}.
\]

This means that our method is effective for recursive functions where the
number of sub-problems is greater than the problem size reduction factor.  This
is intuitive since if the problem size reduction factor is equal to or greater
than the number of sub-problems then adding up the total size of all nodes in
levels above a given node will always result in a sum greater than or equal to
the sum for that node's subtree.

By setting $b$ in $\log b / \log a$ equal to $1$ we get a square root
reduction in space. This should be compared with a pebble game for complete
binary graphs that was reported on in \cite{KSS:2016} in which a similar
recursive structure was considered.

\section{Conclusions and outlook}

We considered the problem of optimizing the implementation of integer
arithmetic on a quantum computer. Prior to our work, the state of the art was
that in order to get a subquadratic overall gate count for a reversible
multiplier a quite significant price had to be paid in that $O(n^{\log_2 3})$
qubits of memory were needed. By using pebble games played on the recursion
tree, we find an improved number of ancillas needed for Karatsuba's recursion,
which turns out to be upper bounded by $O(n^{1.427})$, while maintaining the
asymptotic overall gate count of $O(n^{\log_2 3})$ for the number of gates. An
interesting open problem is to apply these ideas to other recursions, which
leads to the question of finding good pebbling strategies for trees of higher
valency. Another open problem relates to the volume of the circuits for integer multiplication, specifically, whether it is possible to reduce the volume asymptotically below $O(n^{1+\log_2 3})$ and whether non-trivial space-time lower bounds for reversible integer multiplication can be shown that improve over the trivial $\Omega(n^2)$ lower bound for the volume.

\section*{Acknowledgments}
The authors would like to thank BIRS for hosting Banff Seminar 16w5029: Quantum Computer Science, during which part of this research was carried out, and the anonymous referees for providing valuable feedback. 



\end{document}